\documentclass{article}  
\usepackage{breckenridge}
\usepackage{graphicx}
\newcommand{ \be }{\begin{equation}}
\newcommand{ \ee }{\end{equation}}
\newcommand{ \bea }{\begin{eqnarray}}
\newcommand{ \eea }{\end{eqnarray}}
\newcommand{ \la }{\langle}
\newcommand{ \ra }{\rangle}

\frompage{000} \topage{000}                                              
\def\rightmark{E-by-E Net Charge Fluctuations}
\def\leftmark{C. Pruneau et al.}
 
\title{Event by event Net Charge Fluctuations}
\authors{{Claude A Pruneau for the STAR Collaboration}
\\[2.812mm] \normalsize Physics and Astronomy Department,\\
Wayne State University\\
Detroit, Michigan, USA}
 
\abstract{We present analyses of event-by-event dynamical net charge fluctuations measured in130 and 200 GeV Au Au collisions  with the STAR detector. The dynamical net charge fluctuations are evaluated using the $\nu_{+-,dyn}$ observable. Dynamical fluctuations measured in Au Au collisions at 130 and 200 GeV are finite, and exceed charge conservation limits. They deviate from a perfect 1/N scaling and provide an indication that the collision dynamics varies with collision centrality. }

\PACS{25.75.Ld}
\keyword{Relativistic Heavy Ions, Event-by-event fluctuations.}
 
\begin{document}
 
\maketitle
\setcounter{page}{1}

\section{Introduction}\label{intro}

Measurements of fluctuations of conserved quantities such as the electrical charge, baryonic, and strangeness numbers emerge as a new tool to probe the final state of relativistic heavy ion collisions. They may be used to estimate the degree of equilibration and criticality of measured systems. As a specific case, we present measurements of dynamical net charge fluctuations performed with the STAR detector for 130 and 200 GeV Au Au collisions.

Measurements of dynamical net charge fluctuations were  performed using the $\nu_{+-,dyn}$ defined as :
\be
\nu_{+-,dyn} = \frac{\la N_+(N_+-1)\ra}{\la N_+\ra^2} +
    \frac{\la N_+(N_+-1)\ra}{\la N_+\ra^2}
    -2  \frac{\la N_+\ra\la N_-\ra}{\la N_+\ra\la N_-\ra}
\ee 
where $\la O \ra$ expresses an event average of the quantity $O$; $N_+$ and $N_-$ are 
respectively the multiplicity of positive and negative charge particles measured in a given kinematic range.  
The properties of $\nu_{+-,dyn}$ were discussed in details in \cite{Pruneau02}. 
It is directly related to integrals of two-particle densities. 
Identically zero for uncorrelated (Poissonian) particle production,  
it can in principle be either positive or negative. 
Negative values are however expected from +- pair correlations resulting 
from charge conservation, resonance production, etc. Values of $\nu_{+-,dyn}$ 
are furthermore expected to exhibit an approximate $1/s$, with $s$ being the 
number of particle production sources in AA collisions. This scaling shall 
translate in a perfect 1/N scaling (N being the total charge particle multiplicity) 
if the collision dynamics is independent of the collision centrality. $\nu_{+-,dyn}$  
is a robust observable and its use is thus preferred to  observables $D$ and $\omega_Q$ 
used in various theoretical works \cite{Pruneau02,Asakawa00,Jeon99,Jeon00,Koch02}.

Preliminary results were reported at other conferences\cite{Pruneau02a}. The 
130 GeV results will be submitted to a journal shortly\cite{Pruneau03}. 

\section{Measurement}\label{measurement} 

The data presented are from minimum-bias samples of Au + Au collisions at $\sqrt{s_{NN}}=130$ and $200$~GeV acquired by the STAR experiment during the first two years of operation of the Relativistic Heavy Ion Collider. Detailed descriptions of the experiment and the Time-Projection-Chamber (TPC) can be found elsewhere~\cite{Thomas99}.
Events were triggered by a coincidence between the
two Zero Degree Calorimeters (ZDCS) located +/- 18 m from the interaction
center and a minimum signal in the Central Trigger Barrel (CTB), which consists
of scintillator slats surrounding the TPC. The analysis was restricted to events produced within $\pm 0.70$ m (130 GeV data analysis) and $\pm 0.25$ m (200 GeV) of the center of the STAR TPC along the beam axis. In this range, the vertex finding efficiency is 100\% for collisions which result in charged particle multiplicities larger than 50 tracks in the TPC acceptance. It decreases to 60\% for events with fewer than 5 tracks from the primary vertex. Samples of respectively 180000 and 700000 minimum-bias events from the 130 and 200 GeV runs were used in this analysis after cuts. The centrality of the collisions is estimated from the total charged particle track multiplicity detected within the TPC in the pseudorapidity range $|\eta|<0.75$. The fluctuations studies presented in this report are obtained for a pseudorapidity range $|\eta|<0.5$.
Good track quality is required by restricting the analysis 
to charge particle tracks producing more than 15 hits within the TPC. 
One additionally requires that more than  50\% 
of the hits be included in the final fit of the track. 

We present, in Figure 1(a), our measurements of the dynamical net charge fluctuations, 
$\nu_{+-,dyn}$, in Au Au collisions at  130 and 200 GeV as a function of the total 
charged particle multiplicity measured in the STAR TPC within the pseudorapidity range 
$|\eta| \leq 0.75$. The values of $\nu_{+-,dyn}$ are finite and negative implying 
charged particle productions are correlated in these Au+Au collisions.  We find the 
130 and 200 GeV results exhibit the same dependence on total charge particle
 multiplicity  and nearly equal magnitude at any given centrality. This suggests the charge
 particle correlation has a rather small beam energy dependence. Note that a similar 
conclusions was achieved based on detailed studies of two-particle correlations performed 
for $p+p$ and $p+\overline{p}$ collisions conducted at FNAL and ISR\cite{Foa75,Whitmore76,Boggild74}. 

\begin{figure}[htb]
\vspace*{1.5cm}
\epsfxsize=\linewidth\epsfysize=7cm\epsfbox{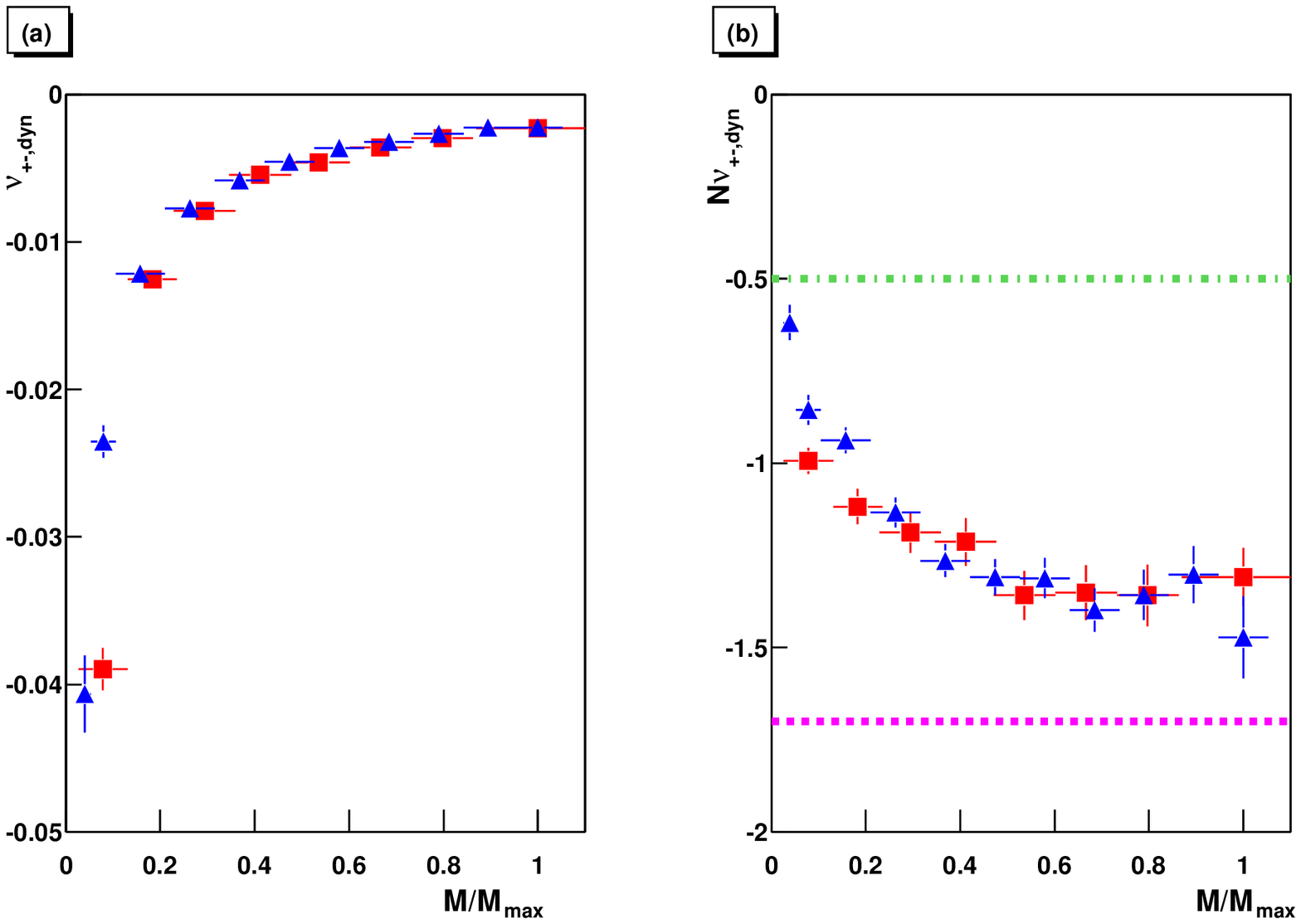}
\vspace*{-1.cm}
\caption[]{(a) Dynamical net charge fluctuations $\nu_{+-,dyn}$ measured for $|\eta|<0.5$ vs relative charged particle multiplicity in  $|\eta|<0.75$: 130 GeV data (square) and 200 GeV data(triangle).\\
(b)Scaled dynamical net charge fluctuations $N \nu_{+-,dyn}$ measured for $|\eta|<0.5$ vs relative charged particle multiplicity in  $|\eta|<0.75$. The dot-dash line indicates the charge conservatin value whereas the dash line
corresponds to a thermalized resonance gas. }
\label{fig1}
\end{figure}

We first consider the fraction of the dynamical net charge fluctuations arising from charge conservation. From ref.\cite{Pruneau02}, we determine the charge conservation (c.c.) contribution to amount to $\nu_{c.c.}= -4/N$ with N the {\em total} produced charge 
particle multiplicity. 

We use the measurement of total charged particle multiplicity  $4200 \pm 470$ in the 
rapidity range $|\eta|<5.6$ reported by the PHOBOS collaboration as an  
estimate of the total charge production at $\sqrt{s_{nn}}=130$ GeV.  We further estimate
the total production at 200 GeV to be larger by a factor of order $1.14\pm 0.05$
based on the ratio of multiplicities in the rapidity range $|\eta|<1$ also
reported by PHOBOS\cite{Phobos02}. 
We thus expect the charge conservation contributions to be of 
order $-0.00095 \pm 0.00011$ and $-0.00084 \pm 0.00010$ at 130 and 200 GeV 
respectively.
 The $\nu_{+-,dyn}$ values measured for 5\% 
central collisions amount to $-0.00236 \pm 0.00006 (stat) \pm 0.00005 (syst)$ and
 $-0.00225 \pm 0.000017 (stat) \pm 0.00005 (syst)$ respectively. The charge conservation contributions thus represent
 only 40\% and 37\% respectively of the observed 
correlations. 
We next compare the strength of the measured correlation to the maximum value one can expect for maximally correlated particle production. This situation arises when 
$N_+ = N_-$ event by event. The maximum value of $\nu_{+-,dyn}$ amounts to $-4/N_{\eta}$ where $ N_{\eta}$  is the total charged particle multiplicity in the  measured 
kinematic  range. STAR measured dn/dy = 523 for 5\% central collisions 
at 130 GeV implying
the maximum value of $\nu_{+-,dyn}$ at this energy is of the order of -0.0076. 
The measured values amount to roughly 30\% of this maximal value. Particle production correlations are thus indeed large in comparison to both the charge conservation 
contribution and the maximum value expected given the observed multiplicities.

 We next discuss the centrality dependence of the dynamical fluctuations.  
The magnitude of $\nu_{+-,dyn}$  is expected to vary as the inverse of the number 
of sub collisions leading to the production of particles. If the number of particles 
produced by such sub-collisions is independent of the collision centrality, 
$\nu_{+-,dyn}$  should exhibit a strict 1/N scaling. We test for such a behavior by
plotting, in Figure 1(b),  $N \nu_{+-,dyn}$  as a function of the total relative charged particle multiplicity M detected in the pseudorapidity interval $|\eta|<0.75$. Both $N$ and $\nu_{+-,dyn}$  are measured in the interval $|\eta|<0.50$. Measured multiplicities $N$ are corrected for finite detection efficiency.
 The efficiency is largest for small TPC occupancies and decreases monotonically by ~10\% at the largest multiplicities. The 130 GeV (square) and the 200 GeV (triangle) data exhibit similar collision centrality dependence. The 200 GeV result, based on a larger event sample, confirms the result
 reported for 130 GeV data in \cite{Pruneau03} that the 1/N scaling is violated.

The magnitude of the measured $N \nu_{+-,dyn}$ increases monotonically by a factor of 
1.5-2 from peripheral collisions to central collisions. By contrast, simulations  with 
HIJING, shown in Figure 2, exhibit a signal independent of the collision centrality. This 
behavior is expected because HIJING essentially models collisions as a  superposition of 
independent nucleon-nucleon interactions. UrQMD simulations shown in Figure 3, on the other hand, include 
re-scattering effects and exhibit a clear centrality dependence\cite{Abdel03}. One finds however the re-scattering
effects included in UrQMD  cause a reduction of the magnitude $N\nu_{+-,dyn}$ for 
central collisions rather than an increase as seen in the experimental data. 

\begin{figure}[htb]
\mbox{
\begin{minipage}{0.48\linewidth}
\begin{center}
\epsfxsize=\linewidth\epsfbox{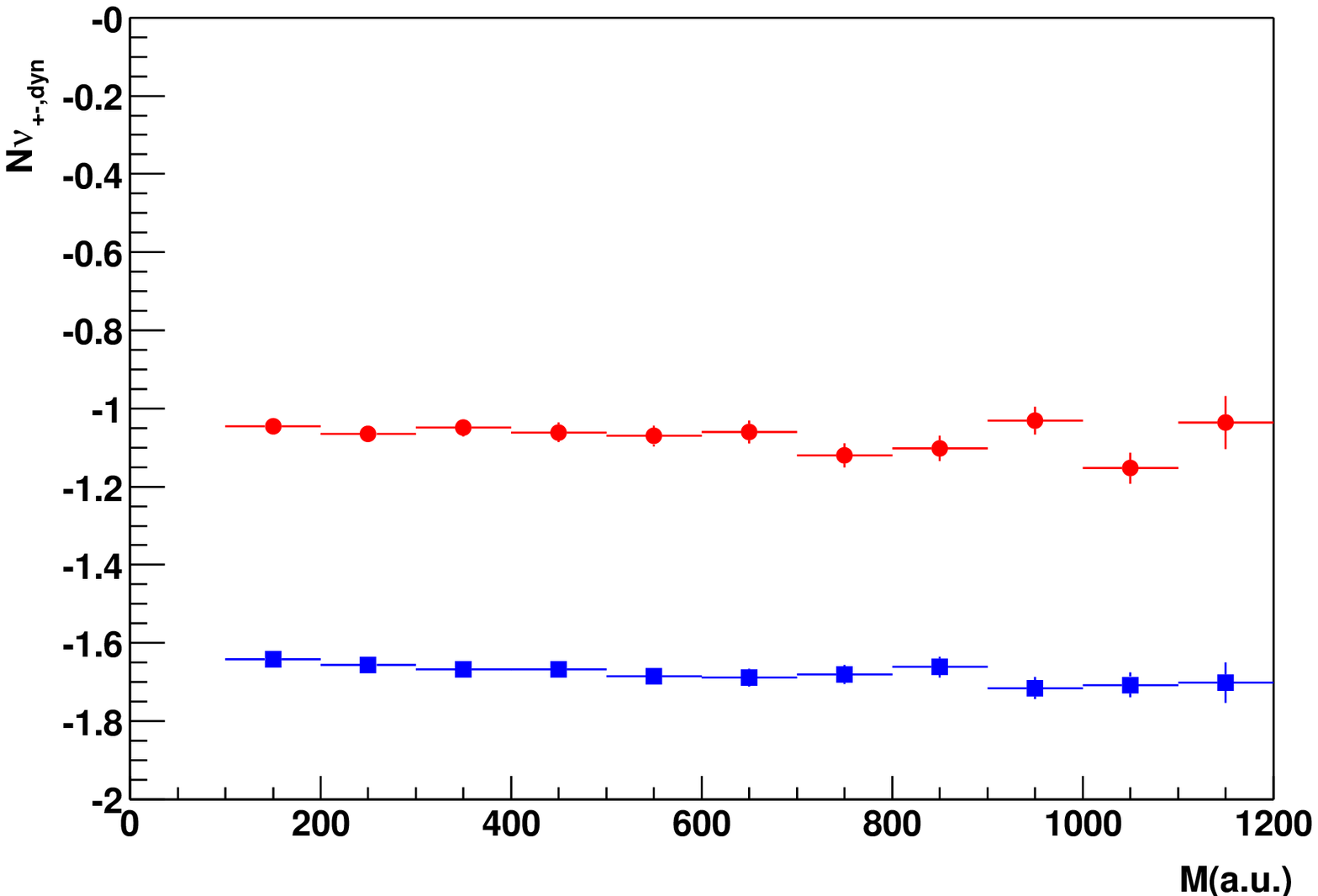}
\caption[]{Calculation of $N\nu_{+-,dyn}$ vs $M(a.u.)$ based on 
HIJING simulations of Au + Au collisions at $\sqrt{s_{NN}}=130$ GeV. Data plotted are 
for $|\eta|<0.5$(circle), and $1.0$(square).}
\label{fig2}
\end{center}
\end{minipage}

\begin{minipage}{0.48\linewidth}
\begin{center}
\epsfxsize=\linewidth\epsfbox{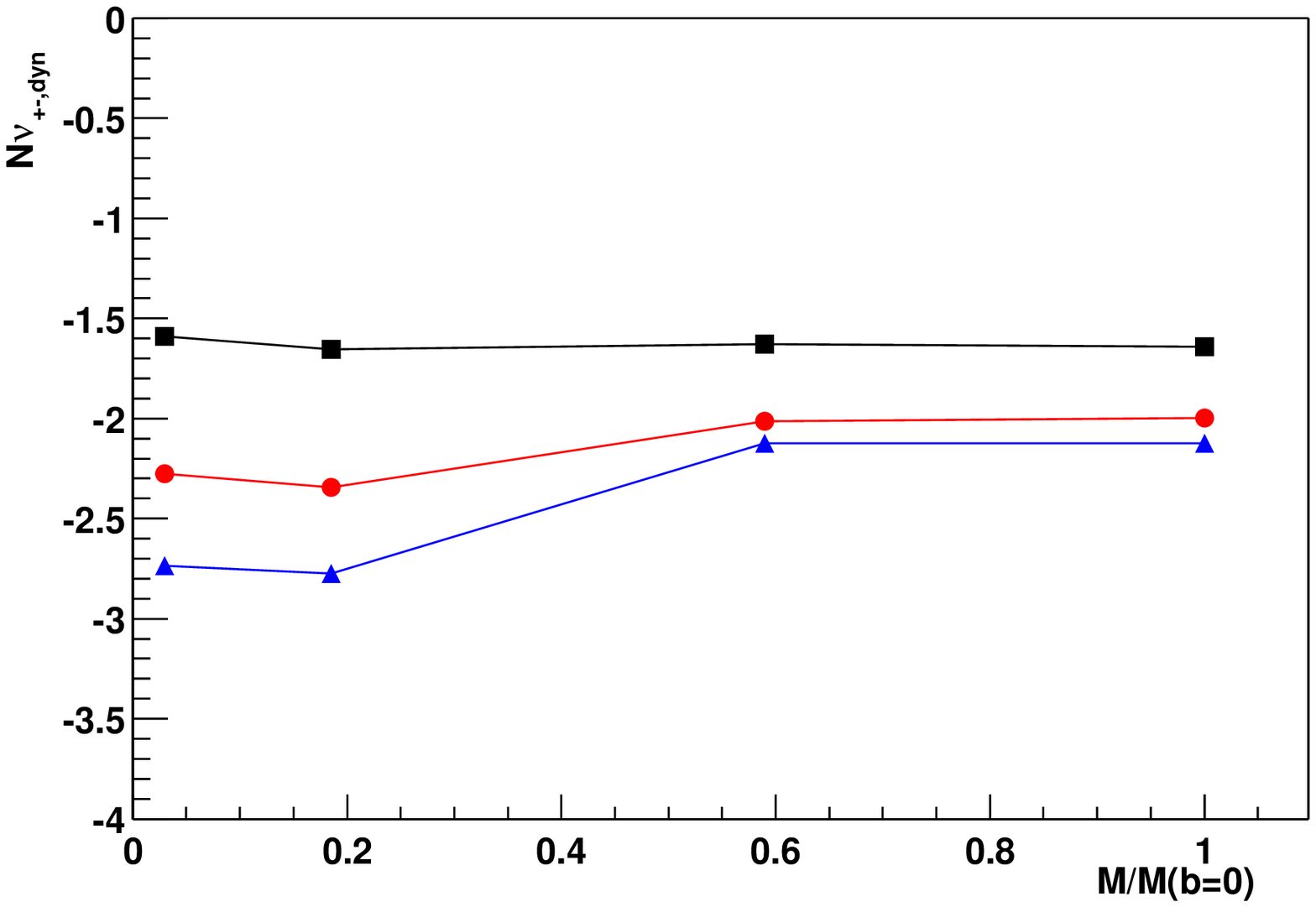}
\caption[]{Calculation of $N\nu_{+-,dyn}$ vs $M/M(b=0)$ based on 
UrQMD simulations of Au + Au collisions at $\sqrt{s_{NN}}=200$ GeV. Data plotted are 
for $|\eta|<0.8$(square), $1.4$(circle), and $2.0$(triangle).}
\label{fig3}
\end{center}
\end{minipage}
}
\end{figure}
We investigate the origin of $N \nu_{+-,dyn}$  measured centrality 
dependence by comparing our data to predictions based on thermal models \cite{Asakawa00,Jeon99,Jeon00,Koch02}. 
To this end, we express our measurement of $\nu_{+-,dyn}$ 
in the range $|\eta|\leq 0.5$ in terms of the $D$ variable 
introduced in \cite{Jeon00}, using
\be
D=4+N\nu_{+-,dyn}
\ee
 valid for $N_+\approx N_-$\cite{Pruneau02}. 
We find using data shown in Fig. 1(b)
 that $D$ decreases from $\approx 3.4$ (statistical error only) 
for the most peripheral 
collisions measured to $2.6\pm 0.1$ in central collisions. 
A comparison to thermal model predictions 
requires the data to be corrected for charge conservation effects. 
One must subtract the charge conservation contribution which 
amounts to $\Delta D = -0.00095\times 526=-0.50 \pm 0.06$. 
The corrected values of $D$ range from $3.9\pm 0.1$ in most peripheral collisions
to $3.0\pm 0.1$ in most central collisions.
The large value measured in peripheral collisions indicates the net
charge fluctuations are essentially Poissonian whereas a sizable correlation is
measured in central collisions. 
According to discussions of Refs.\cite{Asakawa00,Jeon99,Jeon00,Koch02}, 
values measured in central collisions approach that ($D \approx 2.8$) expected for a resonance. They 
are however significantly larger than expected in the above referenced work \cite{Jeon99,Jeon00,Koch02}
for a quark-gluon gas undergoing fast hadronization and freeze-out ($D \approx 1$).
 It is not possible to draw a firm conclusion concerning the existence or non-existence 
of a deconfined phase during the collisions from these results since, 
as the above authors have pointed out, incomplete thermalization could lead 
to larger fluctuations than expected for a QGP. The observable centrality dependence might nonetheless be 
interpreted as suggesting an increased degree of thermalization is reached in most central
collisions relative to more peripheral collisions.

\section{Conclusions}\label{concl}

We have measured event-by-event net charge dynamical fluctuations 
for inclusive non-identified charged particles in Au+Au  collisions at  
$\sqrt{s_{NN}}=130$ GeV and $200$ GeV. Dynamical fluctuations are finite and exceed by nearly a 
factor of two expectations based on charge conservation.  
The violation of the 1/N scaling first observed with the 130 GeV dataset is seen 
and confirmed with better statistical accuracy by data of the 200 GeV dataset. 
Comparison of our measurement with thermal model predictions \cite{Jeon99,Jeon00,Koch02} 
indicate fluctuations, in central collisions, are at a level 
that might be expected if the Au+Au system behaved like a (thermalized) resonance gas. 
The observed collision centrality dependence thus suggests
that an increased degree of thermalization takes place in central collisions 
relative to that achieved in the most peripheral collisions measured in this worked. 
Note finally that although the size of the measured fluctuations is significantly 
larger than expected by Koch {\em et al.} for a quark-gluon gas, limitations of the 
model used prevent a definitive conclusion on the existence or non-existence of a 
quark-gluon plasma phase based on current results.

{\bf Acknowledgment}

The author thanks S.~Gavin and M.~Abdel Aziz for supplying HIJING and UrQMD simulation data.

\vfill\eject
\end{document}